\pdfoutput=1
\documentclass{article}
\usepackage[utf8]{inputenc}
\usepackage{amssymb,latexsym, amsmath, amscd, array, titlesec,bigints,enumitem,fancybox, amsthm, bbold,MnSymbol,verbatim, array}
\usepackage[margin=1.25in]{geometry}
\usepackage{booktabs}
\usepackage{multirow}
\usepackage[export]{adjustbox}[2011/08/13]
\usepackage{float}
\usepackage[dvipsnames]{xcolor}
\definecolor{sage}{RGB}{0,204,153}
\usepackage{graphicx}
\usepackage{microtype}
\usepackage[colorlinks=false, pdfborder={0 0 0}]{hyperref} 
\usepackage{booktabs}
\usepackage{fancybox}
\usepackage{tikz-cd}
\usetikzlibrary{shadows}
\usepackage[framemethod=TikZ]{mdframed}
\usepackage{lipsum}
\usepackage{dsfont}
\usepackage{bbm}
\usepackage{hyperref}
\usepackage{bold-extra}
\usepackage{pstricks-add}
\usepackage{appendix}
\usepackage{float}
\usepackage{pgfplots}
\usepackage[numbers]{natbib}
	\graphicspath{ {downloads/} }
\usepackage{draftwatermark}
\SetWatermarkText{DRAFT}
\SetWatermarkScale{3}
	
\mdfdefinestyle{MyFrame}{%
    linecolor=sage,
    roundcorner=8pt,
    outerlinewidth=2pt,
    innertopmargin=\baselineskip,
    innerbottommargin=\baselineskip,
    innerrightmargin=20pt,
    innerleftmargin=20pt,
    shadow=true,}

\newtheorem{thm}{Theorem}[section]
\newtheorem{theorem}[thm]{Theorem}

\newtheorem{prop}[thm]{Proposition}

\newlist{inline}{enumerate*}{5}
\setlist[inline]{label=(\arabic*)}

\pgfplotsset{width=10cm,compat=1.9}
\usepgfplotslibrary{fillbetween}
\def\cdf(#1)(#2)(#3){0.5*(1+(erf((#1-#2)/(#3*sqrt(2)))))}
\pgfmathdeclarefunction{gauss}{3}{%
  \pgfmathparse{1/(#3*sqrt(2*pi))*exp(-((#1-#2)^2)/(2*#3^2))}%
}

\title{\textsc{Modern Asset Theory}\\
\large \textsc{A Framework for Successful Active Management}}
\author{Corry Bedwell and Ryan Guttridge}
\date{}
\begin{document}
\maketitle
\begin{abstract}
Active management is a term that has many meanings and we have found the defining characteristics needed for success as an “active manager” elusive within the literature. In this paper we offer a set of criteria that defines an active manager and his success. In order to facilitate this, we introduce several definitions, which lead to a logically coherent evaluation framework. We expand on the definitions of these six key concepts:
    \begin{inline}
        \item Introduce a specific concept of Ruin,
        \item Assets,
        \item Risk
        \item Discount rate,
        \item Margin of safety, and
        \item Optimization.
    \end{inline}
    
Through these definitions a strong defense of active management emerges. Furthermore, to the extent one chooses to limit the definitions we offer, our framework reduces to that of Modern Portfolio Theory. Overall, we have aimed to construct a robust expansion of a framework for active management, one that has been found wanting in the current literature.    
\end{abstract}
\section{Executive Summary}
The goal of this paper is to lay out the criteria which define successful active management. 

For active managers to be consistently successful, they need to focus exclusively on the single non-complex dynamic system with which they contend, the “asset”. We find the term asset, although generally accepted, not well defined. To rectify this, we define here the different types of assets a manager can choose from. This expands the mathematical description of assets to include the concepts of dynamics and uncertainty. In other words, each type of asset is a system. Some assets are dynamic, and some are not. 

Active managers need to define an independent concept of value, separate from extrinsic valuations from the interactions of the systems identified above. Finding assets that can intrinsically earn a client's minimum required Normalized Rate of Return, or NRR, is compulsory.  Once identified the Normalized Rate of Return is used to select assets which on a normalized basis earn the required return intrinsically. As a result, for a dynamic asset, cash flow variance, and changes in the normalized operating structure are the critical risk investors face in owning a particular asset. Active managers must not only select assets that will earn the appropriate Normalized Rate of Return but also monitor those assets for that level of intrinsic performance. 

Changes in the valuation resulting from the interaction of complex adaptive systems (Market, and Economy), and assets, are not necessarily a threat to investors earning the NRR overtime. For any given Normalized Rate of Return, a Market price is a valuation of the amount of cash flow the asset will produce. Thus, the price is an estimate of how an asset will perform over time. The difference between this valuation and that implied by the normalized cash flow valued at the NRR, is the “margin of safety”. 

Margin of safety then should be used to optimize position sizes. Active optimization then is the ongoing responsibility of the manager. 

Risk, therefore, is best characterized as a multivariate function with components tied to the asset, the owner of the capital, and price variance. For the owner of the capital, not earning the necessary return is disastrous. The risk the asset will not earn that return is a function of its normalized cash flow and its variance of cash flows. The expected return is a function of the purchase price relative to the present value of the normalized cash flows. 

The active manager needs to incorporate all these concepts into the analytics. This means time and a formal notion of randomness must replace point-estimates in their analytical framework. 

The six actions an Active Manager must do in order to add significant value: 

\begin{enumerate}
    \item Identification of the correct Normalized Rate of Return, or NRR
    \item Identification of an investment that will intrinsically earn the NRR
    \item 3.	Monitoring the investments to assure they continue to intrinsically earn that Normalized return and the variance of the normalized cash flow
    \item Optimizing the portfolio according to the Kelly Criterion as the valuations in the market change
    \item Limit the invested amount to avoid ruin as defined by the Kelly Criterion and allows for additional purchases as edge increases
    \item Optimize overall margin of safety relative to return variance 
\end{enumerate}
\section{Literature Review}
In the 18th century Daniel Bernoulli developed the use of the concave function log(W) as a representation for the utility of wealth. Historically Bernulli introduced this utility function attempting to solve the St. Petersburg paradox.

Edward Thorp re-introduced the Kelly criterion in investment as it maximized log(W), i.e. the utility function introduced by Bernouilli. MacLean, Ziemba, and Thorp have extended and refined the use of the Kelly Criterion as well as many others which has been anthologized in \cite{maclean_thorp_ziemba_kelly}. The novelty in our work is to give a formal definition of edge that relates the Kelly criterion to the classical notion of margin of safety. 

In 2005, Nick French and Laura Gabrielli constructed a valuation method for real-estate assets using Monte-Carlo techniques: modeling discrete functions viewed as point-estimates of a random variable. We extend this method (independently discovered) to the class of securities. Further we tie the Monte-Carlo analysis together with margin of safety, unifying optimization, present-value consumption models, and statistical machinery in a single framework through a formalized classical heuristic. 

Current research by Ole Peters and Murray Gell-Mann focuses on dynamical systems and stochastic utility functions, where the viewpoint shifts and looks at time averages as opposed to ensemble averages. Future work in will be done to incorporate these developments into our framework. 
\section{Modern Asset Theory (MAT)}
This is the first paper in a series. It's goal and the ones that follow, is to demonstrate how different schools of money management persist while apparently being contradictory. We chose to focus on active management vs passive management and when active management can be effective. It is argued, that instead of being in opposition to one another, the two investment approaches persist because they are both functionally and mathematically correct.  However, because of a difference in their fundamental definitions and overall objectives, the systems produce radically different optimal strategies. As we will highlight, the key difference is how the two groups approach risk. From a passive perspective return variance and its derivative price variance are to be avoided or minimized. In contrast, good active managers buy cheap and sell high. This analysis will walk the reader through a set of steps an active manager must follow to capitalize on price variance successfully. From this process, a clear case for the benefits of active management emerges. Additionally, this framework can be reduced to describe how financial professionals work with their clients, either institutional or retail. 	

\begin{quote}
    “After all, if a moody fellow with a farm bordering my property yelled out a price every day to me at which he would either buy my farm or sell me his – and those prices varied widely over short periods of time depending on his mental state – how in the world could I be other than benefited by his erratic behavior?” –Warren Buffett, 2013 letter to shareholders  
\end{quote}

Perhaps the greatest active manager in history is Warren Buffett. In his 2013 letter, he walks the reader through his analytic process. While his comments intuitively describe the benefits of active management, they are packed with nuance and require explanation. In consequence, we use quotes from Buffett to introduce the critical concepts embedded in the MAT approach and then offer a technical explanation of each in the following sections of this paper.  
\section{Asset Classification}
\begin{quote}
    “Owners of stocks, however, too often let the capricious and often irrational behavior of their fellow owners cause them to behave irrationally as well. Because there is so much chatter about markets, the economy, interest rates, price behavior of stocks, etc., some investors believe it is important to listen to pundits – and, worse yet, important to consider acting upon their comments.”
\end{quote}

The outcomes we see in the world are produced by the interaction of complex systems, which we mean complex adaptive system in the formal sense.  As a result, it is necessary not only to identify the independent systems with which investors contend but also to determine which systems are necessary to focus on and which are not. However, first, let us briefly define a complex adaptive system. 

These systems are both complex and adaptive. Thus, they are dynamical systems able to change their structures, interactions, and consequently their dynamics as they evolve. Since they are a function of time, they cannot be analyzed deterministically. Analyzing them requires the use of statistics and their proper distributions to understand how the system will propagate over time \cite{thurner_Klimek_Hanel_complex_2018}.

When making an investment in a publicly traded asset, at least two separate systems must be identified:
\begin{enumerate}
    \item The Economy
    \item The Market
\end{enumerate}

In his 2013 shareholder letter, Buffett tells the story of his purchase of a farm, highlighting the outcome of the interaction between the two systems identified above.

\begin{quote}
    “From 1973 to 1981, the Midwest experienced an explosion in farm prices, caused by a widespread belief that runaway inflation was coming and fueled by the lending policies of small rural banks. Then the bubble burst, bringing price declines of 50\%...”
\end{quote}

The event described above is the interaction between the real estate market, and the economy, systems 1 and 2. The interaction between these two systems creates a certain extrinsic valuation in the real estate market, which is the price of Buffett’s farm.  

\begin{quote}
    “Focus on the future productivity of the asset you are considering. If you don’t feel comfortable making a rough estimate of the asset’s future earnings, just forget it and move on”
\end{quote}

The term "asset" requires further clarification.  Some assets mature, with defined terms such as a bond. Then there are assets, such as Buffett's farm, that never mature and whose cash flow from year to year is volatile and not guaranteed. The former has an end date, and the latter propagates through time. In a complex market and an economy, it is critical to distinguish what kind of asset is in question and then use an analytical approach the deals with time and randomness correctly. These facts give us three general categories of "assets."

\begin{enumerate}
    \item Cash – store of time
    \item  Discrete Assets – Uses of cash that produces a series of cash flows over a specified term, at specified intervals, with specified levels of certainty. The cash flows can be modeled deterministically because it is a deterministic system. An example of a discrete asset is a bond held to maturity.
    \item Dynamic Assets – Uses of cash that propagate through time, and change. They have indefinite life’s while generating variable levels of cash flow for their owners.  Cash flows are multivariate functions and it is a dynamic system. An example of a Dynamic Asset is a publicly traded equity of a company. 
\end{enumerate}

\begin{quote}
    “From these estimates, I calculated the normalized return from the farm to then be about 10\%.”
\end{quote}

Since a dynamic asset is an ongoing system described by random variables, a discrete framework cannot adequately describe it. For example, think of how medical doctors describe your biological functions. They do this with averages, looking for deviations from those averages as signals about your overall health. Similarly, a dynamic asset (read stock or business) should be treated as a system whose operations vary around a set of averages through its lifetime.  The assets cash flow can be defined in terms of a normalized level of operations, and its corresponding variance \footnote{Normalized operating cash flow is defined Normalized Operating Cash Flow as the cash flow net of all average normalized on-going cost and investments. In other words, the average level of cash flow available to equity holders on an ongoing basis.} This normalized operating cash flow is what investors can generally expect to receive during the holding period  (Appendix \ref{AppVal})
\section{Dynamic Valuation: Normalized Rate of Return}

\begin{quote}
    “It should be an enormous advantage for investors in stocks to have those wildly fluctuating valuations placed on their holdings…I could either sell to him or just go on farming.”
\end{quote}
For an active manager to be successful it is clear she must develop an intrinsic view of value. We now introduce the concept of a normalized rate of return. This is distinct from the market multiple. Normalized Rate of Return is specific to the owner or steward of the capital invested in the asset. From this perspective, it is defined by the basic relationship between scarcity and choice \cite{Buchanan_2017}. The notion of Normalized Rate of Return plays a crucial part in attempts to ensure that investors scarce resources are used efficiently \cite{Economics_A-Z_terms_beginning_with_O}. In this case, it is the cost of forgoing other investment opportunities or current consumption. In general, it is the return required to provide a sufficient level of utility to warrant foregoing present consumption. In the formal framework described below, it can be thought of as the multiple an investor is willing to pay for a normalized level of free cash flow. We call this the Normalized Rate of Return Multiple (NRRM), see Appendix  \ref{AppVal}.
\section{Market Price}

\begin{quote}
    “Forming macro opinions or listening to the macro or market predictions of others is a waste of time. Indeed, it is dangerous because it may blur your vision of the facts that are truly important.”
\end{quote}

From a MAT standpoint, the analysis must focus solely on the essential system — the operations of the dynamic asset in question. and its intrinsic properties. Market price (i.e. valuation in the market) is a concern only when the trader is evaluating if a purchase is warranted. and is extrinsic.  No matter its level, a price is a valuation. Given a certain discount rate, it is a spot estimate of a constant operational level of the dynamic asset i.e. a multiple of the assets cash flow. To the extent the market spot estimate reflects an unreasonably high or low variation from its operational averages, it is not sustainable. From this perspective, the price at which Buffett is being offered the farm implies an absurdly low spot estimate of how the farm will run on an on-going basis  (Appendix \ref{AppMoS}).
\section{Market Multiple: Margin of Safety}

The "Market Multiple” defines the “terms of the deal” on offer in the Market (the normalized return multiple relative to the price). With a proper understanding of the NRR multiple, the investor can shop the marketplace for investment opportunities that offer the best relative deal on a cashflow risk-adjusted basis. From this perspective, given a set of opportunities with similar cash flow variances, the definition of Ben Graham’s famous concept of margin of safety becomes:

$$\textrm{Margin of Safety} = \frac{\textrm{NRR Multiple} - \textrm{Market Multiple}}{\textrm{NRR Multiple}} = 1 - \frac{\textrm{Market Multiple}}{\textrm{NRR Multiple}}$$
For a full break down of the significance of the Margin of safety see appendices \ref{AppMoS} and \ref{GB}
\section{Valuation and Price Variance}

\begin{quote}
        “\dots if a moody fellow with a farm bordering my property yelled out a price every day to me at which he would either buy my farm or sell me his…If his daily shout-out was ridiculously low, and I had some spare cash, I would buy his farm. If the number he yelled was absurdly high, I could either sell to him or just go on farming.“
    \end{quote}
    
Because the investor is intrinsically earning his Normalized Rate of Return the current valuation doesn't immediately matter. However, as an inductive premise, the higher the discount from the NRR multiple the greater “margin of safety” embedded in the valuation.  Evidently, the higher the discount, the more likely he is to earn the NRR overtime. Price variance, then, provides an opportunity to increase the size the of the investment, driving down an investors cost basis while asymmetrically increasing the potential overall rate of return by increasing the percent of ownership at a more substantial discount the to the NRRM. 
\section{Risk}

\begin{quote}
    “…I thought only of what the properties would produce and cared not at all about their daily valuations.”
\end{quote}

Since returns over time are a direct function of cash flow, and cash flow is the product of normalized future operations, the risk to that return must be a permanent change in the assets normalized operations. The “risk to the return” needs to be separated from the risk “in the return”. The former is the sustainability of the normalized cash flows. The latter is the actual variance in the normalized cash flows produced by the random variables that define the dynamic asset. Since the risk investors face is not earning their Normalized return intrinsically over time, changes in the distribution of the assets cash flow is the proper way to measure/think about risk. These risks are separate from the inevitable changes in valuation resulting from the on-going interaction between the complex adaptive systems identified above.    
\section{Exit Strategy}

\begin{quote}
    “There would, of course, be the occasional bad crop and prices would sometimes disappoint. But so what? There would be some unusually good years as well, and I would never be under any pressure to sell the property.”
\end{quote}

By monitoring the dynamic-assets performance over time, one can confirm it is operating as expected. This process culminates in a disprovable argument. To maintain a position an investor confirms his estimates of NRRM continue to hold. If evidence emerges that the NRRM has changed, then the position must be re-evaluated and possibly sold. In other words, have the expected parameters of the distribution of cash flow changed (See Appendix  \ref{AppKell})?
\section{Optimization: Price Variance Becomes a Friend}

\begin{quote}
    “It should be an enormous advantage for investors in stocks to have those wildly fluctuating valuations placed on their holdings – and for some investors, it is.”
\end{quote}

This process makes price variance a friend, from two perspectives. First, large movements downward are useful as they pull the market price a long way below that justified by the NRR multiple. This creates a larger edge --- increasing the size of the position demanded by Kelly, providing an opportunity to concentrate positions and driving down the cost basis of the investment. The mean reversion towards the NRRM creates asymmetric returns as a result. The lower the market valuation compared to that justified by the NRR multiple, the higher the percentage of a trader’s portfolio should be allocated to the asset.  Since price volatility increases over time, large changes in market valuation become more likely the longer a position is held.  This makes time an ally in the MAT framework, because the farther below the MM as compared to NRRM, an increase in valuation becomes more likely overtime.    
\section{Optimization: Correlation still matters}

It has been well established, that given the ability to calculate an “edge”, Kelly optimization is the most effective at maximizing the log of the growth in wealth. However, Kelly leads to very high concentrations in few positions. While the geometric return may be optimized overtime, the ability of the trader to maintain the positions long enough for this to be recognized isn’t clear. Also, should the trader’s estimates of normalize cash flow levels prove incorrect, excessive concentration in similar securities leads to Ruin and hence return correlations must be considered.

Price variance is an essential tool in MAT. The ideal optimization would be to capture the most substantial margin of safety possible, with the lowest correlation. We call this relationship the Guttridge-Bedwell ratio, and show in Appendix \ref{GB}, maximizing this ratio provides a classically efficient solution. This final step addresses the three sources of risk identified above. 
\section{Ruin}
\begin{quote}
    “We will buy the stock (or business) if it sells at a reasonable price in relation to the bottom boundary of our estimate.”
\end{quote}

Ruin for the active manager is not earning the Normalized Rate of Return over the specified time period. It is important to realize that this is independent of the market valuation. Kelly optimization requires never making a time-limited bet (read periodic) on the realization of the Normalized Rate of Return. To the extent, capital has a very high near-term cost its Normalized Rate of Return can be thought of as being infinite, and it should never be invested. Ruin is also a separate concept from Risk. Risk is specific to the dynamic assets being employed to earn the NRR overtime. Ruin is what the manager faces by never earning or not earning for sufficiently long enough time the NRR.  Ruin by definition is therefore a function of the NRR and time. 
\section{Active Management}
The active manager role adds significant value from this perspective for five operational reasons:
\begin{enumerate}
    \item Identification of the correct Normalized Rate of Return, or NRR
    \item Identification of investment that will intrinsically earn the NRR
    \item Monitoring the investments to assure they continue to intrinsically earn that Normalized return
    \item Optimizing the portfolio according to the Kelly Criterion as the valuations in the market change
    \item Limit the invested amount to avoid ruin allowing for additional purchases as edge increases
\end{enumerate}

\begin{quote}
    “If you instead focus on the prospective price change of a contemplated purchase, you are speculating.”
\end{quote}
This analytical framework allows for a clear definition of the difference between speculation and investing. 

\textit{Investing}: buying an asset to earn an amount in excess of the Normalized Rate of Return dynamically overtime.

\textit{Speculating}: ignore Normalized Rate of Return and hope for price variance in a certain direction in a certain time period. 
\section{Weaknesses of Modern Asset Theory}
The most obvious weakness in this approach is how does one determine the “correct” NRR? Taken to its logical conclusion it seems to imply that all value is subjective, with each manager arriving at a different level of “personal” fair value.  While we think the addition of random variables, and their resulting distributions, make this point far less problematic. We will address this concern in a future paper, by delving deeper into marginal utility theory. That said we certainly agree more work needs to be done on this point. 

Another obvious criticism is the risk of under-performing the market. It is well documented that most active managers cannot consistently outperform the market on a periodic basis. To the extent the periodic relative performance is the only concern investors and active managers face than this is a valid point. Second, this approach obviously requires locating and hiring a manager capable of identifying the correct assets. Given the data on manager performance, this is likely to be a challenging task. However, it should be apparent to the reader we are only arguing the manager must outperform the NRR, not the market and certainly not on a periodic basis. Also, it needs to be recognized that the 5 points we identify describe what every practitioner, financial planner, and active manager already does in practice. It should be further noted that most of these professionals use some version of passive investing to achieve these outcomes for their client’s. Last, this framework can easily be reduced to the generally accepted viewpoint, by merely limiting the definition of risk to relative return variance and setting the NRR equal to the market return.  
\section{Conclusion}
Active management is justified and necessary when the Normalized Rate of Return is inherent to the capital employed. Given the complexity of the economy and Market, the trader should focus on the least complicated system exclusively, the company. The trader must identify a series of assets capable of intrinsically earning the required NRR. The trader must continually monitor the asset to assure that it is performing intrinsically as expected. Given the complexity embedded in the interactions between economic systems, markets, the valuations placed on companies will change. A manager must be sure to optimize the allocations in his portfolio to maximize the geometric return over time relative to the margin of safety with the lowest correlation amongst assets possible.       

This paper also highlights why practitioners rely so heavily on multiples as they implicitly reflect the dynamic nature of the asset in question. As is shown in Appendix \ref{AppVal}, a Discounted Cash Flow (DCF) framework assigns a multiple to a normalized economic structure. It is seen that the current market multiple is a dynamic measure of value, and its position within its historical range a measure of the margin of safety.  

Finally, we show that active management is not necessarily at odds with modern portfolio theory. The difference between the two constituencies is the result of the definitions they use in practice. The scope of these definitions largely determines how traders attempt to fulfill the five roles identified above. Absent the managers ability to fulfill these roles, a passive low-cost strategy should be employed.  
\appendix
\appendixpage
\addappheadtotoc
\section{Present-Value Consumption Model}\label{AppVal}

This Appendix describes the present-value consumption model employed in the text. In the framework laid out in \cite{cochrane_asset_pricing} this amounts to determining a stochastic discount factor, m, and modeling the parameters for the payoffs, $x$ for investment asset. The fundamental equality is
\begin{mdframed}[linecolor=sage]
\begin{equation}\label{PriceEqn}
    p = E(mx). 
\end{equation}
\end{mdframed}
Where $p$ is the price, $m$ the stochastic discount factor, and $x$ is the payoff. As noted in \cite{cochrane_asset_pricing} $x$ is not to be confused with returns. In particular, given a time period $t$, the payoff for the next time period is given by,
\begin{mdframed}[linecolor=sage]
\begin{equation}\label{payoff}
    x_{i+1} = p_{t+1} + d_{t+1}.
\end{equation}
\end{mdframed}
Where, again $p_{t+1}$ is the price at time $t+1$ and $d_{t+1}$ the dividend return for the interim holding period. This also gives another important equality, as by dividing \ref{payoff} by the price at time $t$, one gets the classic formula,
\begin{equation}
    R_{t+1} = \frac{x_{t+1}}{p_t} = \frac{p_{t+1}+d_{t+1}}{p_t}.
\end{equation}

We take in our framework a purposeful shifting between the conceptual interpretations of \ref{PriceEqn}. The valuations take the stochastic discount factor and future payoffs as known and derive a price, i.e. solving for the left-hand side of \ref{PriceEqn}. Then we shift and re-write the equation as $m = E(\frac{x}{m})$ and solve for the stochastic discount factor $m$. In the former we are assuming that the discount rate and payoffs are known, in the latter that the known values are price and payoffs.  

Fix $n$ and let $f_i$, for $i = 1\dots n$, denote a random variables estimating value drivers of the underlying security. For example, we could choose $f_1 =$ COGS, Cost of goods sold. Then for an appropriate choice of $f_1,\dots,f_n$ we define the current valuation of the security as follows.
\begin{equation}\label{valuation}
    V_t = \sum_{j=1}^{\infty}\frac{F[f_1,\dots,f_n](t+j)}{(1+N)^{t+j}},
\end{equation}
where $N$ is the NRR (normalized rate of return) described in the text. The normalized rate of return multiple, NRRM, is $NRRM = \frac{1}{N}$. Note that in this frame work \ref{PriceEqn} becomes,
\begin{equation}
    p_t = \sum_{j=1}^{\infty}m_{t+j}x_{t+j},
\end{equation}
where $m_{t+j} = (1+N)^{-(t+j)}$ and $x_{t+j} = F[f_1,\dots,f_n](t+j)$.

We take as a practicality that $N\in (0,1)$. The theory works without this, however most applications have this requirement built in naturally. We also require two stipulation for the above series.
\begin{enumerate}
    \item Let $F_t$ denote $F[f_1,\dots,f_n](t+j)$ and like wise $N_t = (1+N)^{t+j}$ then we have,
    $$\lim_t \left | \frac{F_{t+1}N_t}{F_tN_{t+1}}\right | < 1$$
    \item $F_t$ is a real valued, continuous function, so that the numerator in the series is again a random variable.
\end{enumerate}
For $f_1(t) = g$, where $g$ is the fixed growth rate of the dividends, $F[f_1](t+j) = D_0(1+f_1)^{t+j}$ with $D_0$ the initial dividend amount, and $(1+N)^{t+j} = (1+r)^{t+j}$ with $r$ the WACC, then \ref{valuation} becomes,
$$V_t = \sum_{j=1}^{\infty}\frac{D_0(1+g)^{t+j}}{(1+r)^{t+j}}.$$
That is \ref{valuation} reduces to the Gordon growth model for valuation. 
\newline\indent The previous example illustrates how the definition of \ref{valuation} generalizes a valuation method currently in practice. The more theoretically satisfying use of \ref{valuation} will be discussed following the framework for the Kelly criterion.
\section{Margin of Safety}\label{AppMoS}
In this appendix we give a formal definition of margin of safety. Conceptually, the margin of safety is the discrepancy between the shifting from $p = E(mx)$ to $m = E(\frac{x}{p})$. On the one hand we derive a price taking the stochastic discount factor and payoffs as given. Then we hold the price and payoffs fixed and compute the derived stochastic discount factor dictated by the market. The normalized difference is the margin of safety.  

Given a valuation we can now define margin of safety.  That is, we are given
    \begin{equation*}
        P_t = \sum_{j=1}^{\infty}\frac{F[f_1,\dots,f_n](t+j)}{(1+N)^{t+j}}
    \end{equation*}
From elementary calculus there is a $c \in (1,2)$, which we will call the growth constant for the valuation, so that 
    \begin{equation*}
        P_t = \sum_{j=1}^{\infty}\frac{c^{t+j}}{(1+N)^{t+j}} = \frac{1}{1-\frac{c}{1+N}}.
    \end{equation*}
Consider now the current market price, $P_t^m$, offered in the market. A staple in the Value investment paradigm is to search for prices in the market that are less than the valuation ("cheap" assets). We therefore assume that $P_t^m < P_t$. Using the same method as above we can write $P_t^m$ as
    \begin{equation*}
        P_t^m = \sum_{j=1}^{\infty}\frac{c^{t+j}}{(1+M)^{t+j}} = \frac{1}{1-\frac{c}{1+M}},
    \end{equation*}
for some $M \in (0,1)$. Since $c$ is fixed $N < M$. Making use of the completeness of the real number, we can prove the important result:
\begin{prop}
For the $M$ above,
$$P_t^m = \sum_{j=1}^{\infty}\frac{c^{t+j}}{(1+M)^{t+j}} = \sum_{j=1}^{\infty}\frac{F[f_1,\dots,f_n](t+j)}{(1+M)^{t+j}}.$$
\end{prop}
With the proposition we can define the market multiple as $PM = \frac{1}{M}$. Finally, this gives us the framework to define the margin of safety.
\begin{mdframed}[linecolor=sage]
\begin{equation}\label{MOS}
    \delta = \textrm{Margin of Safety} := \frac{\frac{1}{N} - \frac{1}{M}}{\frac{1}{N}} = 1 - \frac{N}{M}.
\end{equation}
\end{mdframed}

We collect some facts and leave the calculations to the reader.
\begin{table}[H]
    \centering
    \renewcommand{\arraystretch}{1.5}
    \begin{tabular}{m{10em} m{10em} m{10em}}\toprule
        $P =$ price & $m =$ market multiple & $c =$ growth constant \\\midrule 
        Equality & & Derivation \\\midrule
       \multirow{2}{7em}{$P = \frac{1+M}{1+M-c}$}  & & $\frac{dP}{dc} = \frac{1+M}{(1+M-c)^2}$\\
        & & $\frac{dP}{dM} = \frac{-c}{(1+M-c)^2}$\\
        \cmidrule{1-2}
         \multirow{2}{7em}{$c = \frac{(P-1)(1+M)}{P}$} & & $\frac{dc}{dM} = \frac{P-1}{P}$\\
        & & $\frac{dc}{dP} = \frac{1}{P^2}(M+1)$\\
        \cmidrule{1-2}
        \multirow{2}{7em}{$M = \frac{P}{P-1}c - 1$} & & $\frac{dM}{dc} = \frac{P}{P-1}$\\
        & & $\frac{dM}{dP} = \frac{-1}{(P-1)^2}$\\
        \bottomrule
    \end{tabular}
    \caption{Relation between price, necessary rate of return multiple, and the growth constant.}
    \label{tab:my_label}
\end{table}

The crucial take away from \ref{tab:my_label} is that M and c are \textit{linear} with respect to the other, yet they are \textit{not} with respect to price. This gives the asset manager a new tool for discerning between equally attractive assets. Recall from the literature that the percent margin of safety is defined as 
\begin{equation*}
    S = \frac{P_t - P_t^m}{P_t} = 1-\frac{P_t^m}{P_t}.
\end{equation*}
This gives the percentage of the of the expected margin of safety. That is, for example, if $S = .75$ then the market price is \%25 of the valuation or \%75 less the valuation. This leads to a natural problem with asset selection. Two securities could have the same percent margin of safety. This is where the margin of safety as defined in \ref{MOS} has an advantage. 

With M and c non-linear in price, $\delta$ can give new information not captured in $S$. In particular, one can construct an example with two securities $A_1$ and $A_2$ such that $S_1 = S_2$ and $\delta_1\not = \delta_2$. That is $\delta$ gave a distinction between the assets whereas $S$ failed. We have introduced a non-trivial new definition of margin of safety. 
\section{GB-Ratio: A partial order on assets}\label{GB}
This Appendix outlines the GB-ratio concluding with its relation to a partial order on assets. We define the GB-ratio as the ratio of the margin of safety to the historical price variance, 
\begin{mdframed}[linecolor=sage]
\begin{equation}
    GB-ratio = \frac{\delta}{\sigma}.
\end{equation}
\end{mdframed}

As was discussed in Appendix \ref{AppMoS}, the margin of safety falls short as a partial order on assets. The GB-ratio fills in the gap. The analogy that is useful is that of the relation between the Sharpe ratio and the efficient frontier. The coordinate $(\delta,\sigma)$ in the margin of safety and variance plane yeilds a way to partially order assets. Those assets that maximize the margin of safety for a fixed variance are "efficient". 
\section{Monte-Carlo Analysis}\label{AppMon}

This appendix briefly walks the reader through Monte-Carlo Analysis as a method(s)\footnote{There are technically two methods that are referenced as Monte-Carlo, and the literature uses the one label often without noting which method is being deployed. There is a subtle distinction between the two.} and describes the application of both methods in the setting of asset valuation. The synopsis is 
\subsection{Set Up}\label{MC-setup}
    Let $X$ be a random variable. In this case $\Omega = \mathbb{R}$, and let $f$ be the probability distribution function for $X$. Suppose that $g: \mathbb{R}\to\mathbb{R}$ is continuous. Then, by definition, the expectation of $g$ is given by, 
    \begin{equation}
        E[g(X)] = \int_{\Omega = \mathbb{R}}g(x)f(x)dx.
    \end{equation}
    This leads to the two methods of Monte-carlo simulation (analysis) outlined in the next two subsections.
\subsection{Method One}
    Recall from standard real analysis, the Chebychev inequality:
    \begin{theorem}[Chebychev's Inequality]
\begin{equation*}
    p(\{x\in\mathbb{R}|h(x)\geq \lambda\})\leq \frac{1}{\lambda}\int_{\Omega=\mathbb{R}}h(x).
\end{equation*}
    \end{theorem}
    Which has the corolary for the above set up in \ref{MC-setup},
    \begin{equation}\label{MC1}
        E[g(X)] \simeq \frac{1}{N}\sum_{i=1}^N g(X_i).
    \end{equation}
    Where the $X_i$ are sample trials of $X$. Equation \ref{MC1} gives us the first method, which we will call Monte-Carlo analysis to distinguish the first from the second method. One need only sample a random variable to obtain an approximation for an expected value without the need to know the pdf of $X$. Therefore, in this method, given $\{X_i\}_1^N$ discrete values, one assumes that the $X_i$ are samples of a random variable and makes the expectation estimate using \ref{MC1}. 
\subsection{Method Two}
\subsection{Monte-Carlo and Present-Value}
This subsection tackles the syntesis of method one and two from above to achieve a statistically rigorous Present-Value consumption based model. This subsection also lays down a key component in the Kelly criterion for portfolio optimization. The importance of the Monte-Carlo process to both the pricing and optimization dictated the necessity to draw the distinctions in the two methods and the thoroughness of the overall appendix. 
\section{Kelly Criterion}\label{AppKell}
It has been shown in the literature that the Kelly criterion maximizes the logarithmic utility function of wealth as time tends to infinity. Further, among weighting strategies, the Kelly criterion dominates in the following sense \footnote{
The theorem as stated is lacking the full power found in the literature, however is sufficient for the purposes of the paper.
}:
\begin{theorem}
\begin{enumerate}
    \item Given two portfolio strategies, if strategy $\Lambda^*$ maximizes $\mathbb{E}[\log(W)]$, and strategy $\Lambda$ is any "essentially different" strategy, then 
$$\lim_n \frac{X_n(\Lambda^*)}{\overline{X_n(\Lambda)}}\to \infty$$
\item The expected time for the current capital $X_n$ to reach any fixed preassigned goal $C$ is, asymptotically, least with a $\Lambda^*$ strategy. 
\end{enumerate}
\end{theorem}
The challenge inherent in the Kelly Criterion when applied to non-deterministic probabilities is determining the appropriate approximations for $p$ and $q$ in \ref{wager}. We describe a solution to this problem now. Given $f_1,\dots,f_n$, the random variables characterizing the company fundamentals needed for a discounted cash flow model, we define $\hat{f}_i = \frac{f_i}{R}$. Where $R(t)$ is the total revenue at time $t$. That is, we normalize $f_i$ with respect to its percentage of revenue. Let $g(t) = \frac{R(t) - R(t-1)}{R(t)}$, the percent growth in revenue, and $\overline{g} = \mu(g)$ the historical average of percent revenue growth. Finally, we let $\mu_i = \mu(\hat{f}_i)$ denote the averages of the normalized fundamental factors. With this we can determine an expected future revenue and fundamental random variables via the equality's,
\begin{equation}\label{ff}
    f_i(t) = R(t)\ast(\mu_i)
\end{equation}
and
\begin{equation}\label{fRev}
    R(t) = \frac{R(t-1)}{1 - \overline{g}(t)}.
\end{equation}
Note from \ref{ff} and \ref{fRev}, the future revenue and fundamental random variables can be estimated from historical averages of fundamentals and growth, along with prior rates of revenue. 
\newline\indent Here is where the switch from real valued functions to a set of random variables is more than just an abstraction. We are now in a place to utilize Monte-Carlo analysis. We remind the reader that $\mu_i = \mu(\hat{f}_i)$, and we let $\sigma_i$ be the corresponding standard deviation. Fix an $f_k$ for some $k$ in $1,\dots,n$.
\newline\indent We therefore have a Monte-Carlo simulation running on each of the normalized fundamental factors for the discounted cash flow model. As a result we obtain a distribution of implied prices for the underlying security. We then let $P_t$ denote the average of the distribution and $\sigma_0$ the co-responding standard deviation. 
\newline\indent We can now address the probabilities $p$ and $q$ in the Kelly criterion. In particular, $p$ is a function of the standard deviation of the market price $P_t^m$ in relation to $P_t$. We remind the reader of a few definitions and facts from statistics.  
\newline\indent For a random variable $X$ with normal distribution, mean $0$, and standard deviation $\frac{1}{2}$, the cumulative distribution function is defined as,
\begin{equation}
    \Phi(x) = P(X\leq x) = \frac{1}{\sqrt{2\pi}}\int_{-\infty}^x e^{\frac{-t^2}{2}}dt,
\end{equation}
and for a random variable with mean $\mu$, and standard deviation $\sigma$,

\begin{equation}\label{cdf}
    \Phi_{\mu,\sigma}(x) = \Phi\left(\frac{x-\mu}{\sigma}\right).
\end{equation}
When the context is clear, we will drop the subscript notation in equations like \ref{cdf}. To get the weighting of the security to drop as the standard deviation approaches the current Present-value we make the technical re-normalization's of $Phi$ given in \ref{renorm} below.
\begin{mdframed}[linecolor=sage]
\begin{equation}\label{renorm}
    \psi(x)  = 1-2\Phi(x)
\end{equation}
\end{mdframed}

\begin{figure}[H]\includegraphics[scale=.6,center]{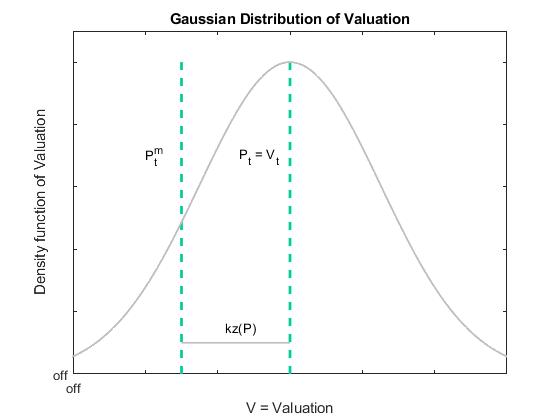}
\label{fig: Gauss Val}
\end{figure}

See the figure \ref{fig: Gauss Val} below for a visual representation of $\Psi(x)$.

The idea then, is to weight securities with price less than the present valuation with higher wagers as a function of what fraction of standard deviations less than the present valuation, and to reduce the position size as the market price gets to the present valuation or higher. Therefore, from the Kelly Criterion, we define
\begin{equation}
    p = p(P_t^m) = \Psi_{P_t,\sigma_0}(P_t^m),
\end{equation}
and by definition,
\begin{equation}
    q =1 - p(P_t^m) = 1-\Psi_{P_t,\sigma_0}(P_t^m).
\end{equation}

Let $P_t^m$ demote the current price in the market. Then we define the edge for the Kelly criterion, $e$, as
$$e:= (P_t)p - (P_t^m)q.$$
Which leads to the optimal wager for the Kelly criterion for one security:
\begin{mdframed}[linecolor=sage]
\begin{equation}\label{wager}
    wager = \frac{(P_t)p - (P_t^m)q}{P_t}.
\end{equation}
\end{mdframed}

A graphical representation of $p$, and therefore $q$, computed above is shown below in figure \ref{fig: ERF box}. We have also highlighted the coordinate $(0,0)$, which corresponds to $p=0$. In this case the portfolio becomes market weighted. \begin{figure}[H]\includegraphics[scale=.6,center]{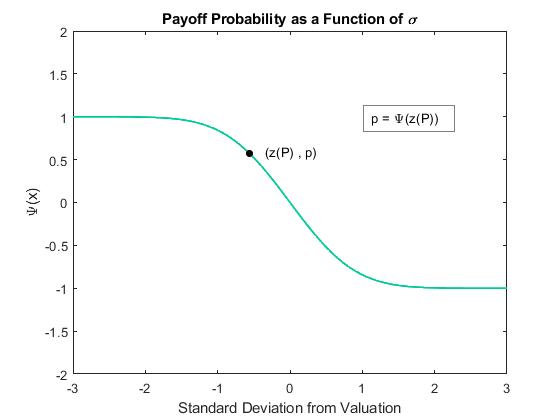}
\label{fig: ERF box}
\end{figure}
\begin{figure}[H]\includegraphics[scale=.6,center]{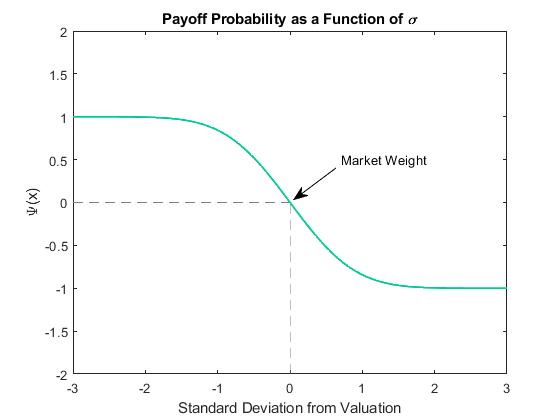}
\label{fig: ERF weight}
\end{figure}

Note that this gives the active manager a clear answer to the question, "When to sell a stock?". For a fixed $P_t$, that is so long as the securities fundamentals remain stable, then a decrease in market price $P_t^m$ gives a larger implied wager according to \ref{wager}. In fact, the position size is reduced if and only if $|(P_t)p - (P_t^m)q|$ tends to zero. This can happen either with a change in the company fundamentals, or a market price that reflects the current valuation.
\nocite{*}
\bibliography{bib}
\bibliographystyle{abbrv}
\end{document}